\documentclass[a4paper]{aipproc}
\DeclareSymbolFont{AMSa}{U}{msa}{m}{n}
\DeclareMathSymbol{\gtrsim}       {\mathrel}{AMSa}{"26} 
\DeclareMathSymbol{\lesssim}      {\mathrel}{AMSa}{"2E}
\layoutstyle{6x9}
\begin{document}

\title [Photoproduction of Heavy Quarkonia] {Photoproduction of Heavy
Quarkonia\protect{\footnote{Talk given by B. J\"ager at \lq\lq
Mesons and Light Nuclei `01\rq\rq, Prague, Czech Republic, July 2001}}}


\author{B. J\"ager}{
  address={Institute of Theoretical Physics, University of 
  Regensburg, \\D-93040 Regensburg, Germany},
  email={barbara.jaeger@physik.uni-regensburg.de},
  thanks={This work was supported by the Paul-Urban-Stipendienstiftung}
}

\iftrue
\author{W. Schweiger}{
  address={Institute of Theoretical Physics, University of 
  Graz, Universit\"atsplatz 5, \\A-8010 Graz, Austria},
  email={wolfgang.schweiger@uni-graz.at},
}

\fi


\begin{abstract}
We investigate the reaction $\gamma p \rightarrow V p$, with $V$
denoting a $\Phi$ or a $J/\Psi$ meson, within the scope of
perturbative QCD, treating the proton as a quark-diquark system.  Our
predictions extrapolate the existing forward differential cross-section
data into the few-GeV momentum-transfer region.  In case of the
$J/\Psi$ reasonable results are only obtained by properly taking into
account its mass in the perturbative calculation of the
hard-scattering amplitude.
\end{abstract}

\date{\today}

\maketitle

\section{Introduction}
Within the last few years exclusive photoproduction of vector mesons
($\rho, \omega, \Phi, J/\Psi$) has attained increasing interest which
has been stimulated by corresponding experimental efforts at DESY and
TJLab.  The investigation of high-energy diffractive photoproduction
at DESY aims at a better understanding of Pomeron phenomenology in
terms of QCD. The 93-031 experiment at TJLab, on the other hand, is
situated in the few-GeV (momentum-transfer) region and tries to shed
some light on the transition from the non-perturbative (vector-meson
dominance) to the perturbative (quark and gluon exchange) production
mechanism.  For heavy-quarkonium channels the perturbative production
mechanism becomes particularly simple.  Whereas $\rho$ and $\omega$
production can proceed via both, quark and gluon exchange, heavy
quarkonia are produced via gluon exchange mainly.  Quark exchange is
power-suppressed due to the small heavy-quark content of the nucleon.

In the present contribution we report on an attempt to describe
photoproduction of $\Phi$ and $J/\Psi$ mesons in the perturbative
regime by means of a modified version of the hard-scattering approach
(HSA), in which the proton is treated as a quark-diquark rather than a
three-quark system.  This perturbative diquark model accounts
effectively for quark-quark correlations inside a baryon and has
already successfully been applied to the investigation of various
exclusive photon-induced reactions
\cite{JaKroSchuSchw93,KroPiSchuSchw93,KroSchuGui96,Ber97,KSPS97,BeSchw00}.
It provides a consistent description of baryon electromagnetic form
factors, Compton scattering off baryons, photoproduction of $K$
mesons, etc., in the range of intermediate momentum transfers
($p_{\perp}^2\gtrsim 3$ GeV$^2$) in the sense that data for these
reactions are reproduced with the same set of model parameters.

\section{The Hard-Scattering Approach with Diquarks}
Within the HSA an exclusive scattering amplitude $M$ can be written as
a convolution integral of a perturbatively calculable hard-scattering
amplitude $T_{\{\lambda\}}$ with distribution amplitudes $\phi_H$,
which contain the bound state dynamics of the involved hadrons $H$.  For
the photoproduction reaction $\gamma p \rightarrow V p$ this integral
takes on the particular form
\begin{equation}\label{Eq-M}
M_{\{\lambda\}}(\tilde{s},\tilde{t})=\int_0^1dx_1dy_1dz_1\phi_{V}^{\dagger}(z_1)
\phi_{p}^{\dagger}(y_1)T_{\{\lambda\}}(x_1,y_1,z_1;\tilde{s},\tilde{t})\phi_p(x_1).
\end{equation}
The distribution amplitudes $\phi_H$ are probability amplitudes for finding the
valence-Fock state in the hadron $H$ with its constituents carrying
certain fractions $x_i, y_i, z_i,\; i=1,2$ of the momentum of their
parent hadron.  In the diquark approach the valence-Fock state of a
baryon is assumed to consist of a quark ($q$) and a diquark ($D$). 
The hard-scattering amplitude $T_{\{\lambda\}}$ is calculated on tree
level in collinear approximation.  The subscript $\{\lambda\}$ of $T$
and $M$ represents a set of possible photon, proton, and vector-meson
helicities.  The analytical results are conveniently expressed in
terms of Mandelstam variables $\tilde{s}, \tilde{t}$, and $\tilde{u}$, which
are obtained by neglecting the proton mass.  All hadron masses,
however, are fully taken into account in flux and phase-space factors.

The model, as applied in
Refs.~\cite{JaKroSchuSchw93,KroPiSchuSchw93,KroSchuGui96,Ber97,
BeSchw00} includes scalar as well as axial-vector diquarks.  Feynman
rules, the quark-diquark distribution amplitude of the proton, and
further details of the diquark model can be found in
Ref.~\cite{JaKroSchuSchw93}.  The numerical values of the model
parameters for the present work are also adopted from that paper.  In
order to describe $\Phi$- and $J/\Psi$-meson photoproduction
simultaneously we have modified the $\Phi$-meson distribution
amplitude proposed by Benayoun and Chernyak~\cite{BenChe90} by
attaching a flavor dependent exponential factor (cf. 
Ref.~\cite{BeSchw00}).

Unlike the pure HSA, in which hadron masses are completely neglected
in the calculation of the hard-scattering amplitude, we include them
in our calculation.  This improves the applicability of the model at
momentum transfers of only a few GeV, where mass effects may still be
important.  Our treatment of mass effects parallels the one described
in detail in Ref.~\cite{BeSchw00}.  In that work an expansion in
powers of $(m_H/\sqrt{\hat{s}})$ at fixed scattering angle has been
performed.  Only leading order and next-to-leading order terms in the
expansion have been kept.  Whereas this procedure gives a reasonable
description of photoproduction of $\Phi$ mesons in the few-GeV
momentum-transfer range, it fails in the case of the heavier $J/\Psi$
mesons.  Therefore we take the $J/\Psi$ mass fully into account,
treat the proton-mass, however, still like in Ref.~\cite{BeSchw00}.

\section{Results and Conclusions}

\begin{figure}
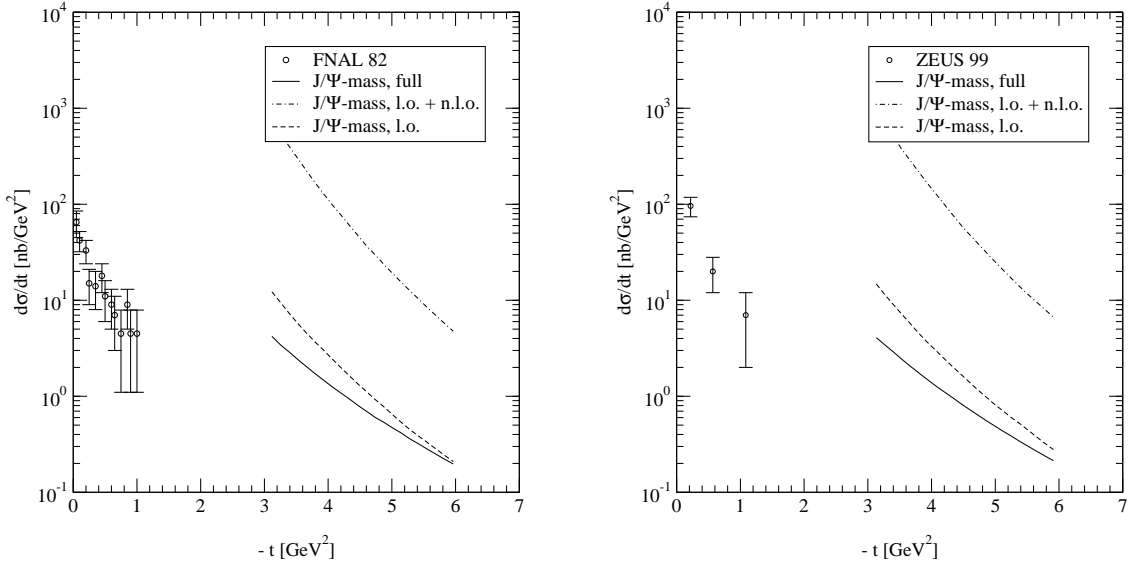
\label{Fig-jpsi}
\hspace*{\fill}\includegraphics[height=.34\textheight,clip=]
{jpsi150.eps}
\hspace*{1cm}\includegraphics[height=.34\textheight,clip=]
{jpsi4708.eps}\hspace*{\fill} \caption{Differential cross section for
$\gamma p\rightarrow J/\Psi p$ versus momentum transfer $-t$ at photon
laboratory energies of 150 GeV (left) and 4708 GeV (right).  Data are
taken from Binkley et al.~{\protect\cite{Bin82}} and Breitweg et
al.~{\protect\cite{Bre00}}, respectively.  Diquark model predictions
are shown for different treatments of the $J/\Psi$ mass: full
inclusion of the $J/\Psi$ mass (solid), leading (dashed), and
next-to-leading order (dash-dotted) of an expansion in
($M_{J/\Psi}/\sqrt{\tilde{s}}$).}
\end{figure}

Photoproduction of $\Phi$ mesons in the forward-scattering domain can
be well described by simple Pomeron phenomenology \cite{DoLan89}.  At
higher momentum transfers a QCD-inspired version of the
Pomeron-exchange model has been proposed by Laget and Mendez-Galain
\cite{LagMe95}, in which the Pomeron is replaced by two (Abelian)
gluons.  If only those graphs are taken into account in which the two
gluons couple to the same quark in the proton, the two-gluon cross
section exhibits a characteristic node due to an interference
of the two different Feynman diagrams contributing to the
photoproduction amplitude.  By additionally considering diagrams, in
which the gluons couple to different quarks in the proton this node is
completely washed out~\cite{Lag00}.  For $|t| \gtrsim 4$ GeV$^2$ the
resulting differential cross section becomes then comparable with the
diquark-model prediction~\cite{BeSchw00}.  This similarity is not
surprising since it is known that the hard-scattering mechanism, i.e.
diagrams without loops, in which all hadronic constituents are
connected via gluon exchange, becomes dominant in the kinematic range
of large $t$ and~$u$.

Difficulties arise if $J/\Psi$ production is considered.  Both,
Pomeron exchange as well as two-gluon exchange overestimate the
differential cross section by at least one order of magnitude.  The
results are improved by arguing that the coupling of the Pomeron (or
the two gluons) to a charmed quark is weaker than to a strange quark. 
If hadron masses are treated like in Ref.~\cite{BeSchw00} similar
difficulties are encountered within the diquark model.  With the
modified treatment of mass effects in which the meson mass is fully
taken into account (for details see, e.g., Ref.~\cite{Jae01}), this
deficiency is, however, cured.

Figure~\ref{Fig-jpsi} shows the diquark-model predictions for
$d\sigma/dt$ at photon energies of 150~GeV and 4708~GeV, respectively,
together with experimental data from FNAL~\cite{Bin82} and
DESY~\cite{Bre00}.  Due to the lack of data in the region of large
transverse momentum transfers (i.e. large $t$ and $u$) a direct
comparison of our predictions with experiment is not possible, since
the perturbatively inspired diquark model is not expected to be
applicable in the kinematic region of $p_{\perp}^2\lesssim 3$ GeV$^2$. 
The diquark-model predicitions with the $J/\Psi$ mass fully included,
however, extrapolate the low-$t$ data in a reasonable way.  By
comparing the leading and next-to-leading-order results of the
expansion in $(M_{J/\Psi}/\sqrt{\tilde{s}})$ it becomes obvious that
such a series expansion converges rather poorly.  A closer inspection
reveals that the expansion coefficients are angular dependent.  The
leading-order terms are only dominant for sufficiently large
deflections.  Towards smaller scattering angles mass-correction terms
become increasingly important.  For FNAL and HERA energies momentum
transfers of a few GeV just mean that we deal with nearly forward
scattering and thus we should refrain from an expansion in the $J/\Psi$
mass.

As one would expect, the effect of taking the meson mass fully into
account is much less pronounced in $\Phi$ production than in $J/\Psi$
production.  For equal values of the mass-expansion parameter
$(M_{V}/\sqrt{\tilde{s}})$ and of Mandelstam $t$ the photon laboratory
energy for $\Phi$ production has to be smaller by about a factor
$(M_{\Phi}/M_{J/\Psi})^2 = 0.11$ and correspondingly the scattering
angle has to be larger, so that the leading-order terms in the mass
expansion become dominant.  But also in $\Phi$ production the full
inclusion of the $\Phi$ mass has a positive effect.  It slightly
improves the angular dependence of the differential cross-section as
compared to the approximate mass treatment of Ref.~\cite{BeSchw00}.  A
full account of diquark-model predicitions for $\Phi$ and $J/\Psi$
photoproduction, which includes also spin observables, can be found in
Ref.~\cite{Jae01}.

To sort out whether perturbative photoproduction mechanisms already
start to dominate in the few-GeV momentum-transfer region precision
data for $|t| \gtrsim 3$~GeV$^2$ are doubtlessly needed.  A severe
test for a perturbative model, like the diquark model, would, however,
be its confrontation with data for polarization observables.  Hard
scattering is closely connected with hadronic helicity conservation. 
Polarization observables could help to reveal whether the inclusion of
constituent-masses and two-quark correlations in terms of diquarks
suffices to model higher-twist and non-perturbative effects, or
whether other (non-perturbative) mechanisms have to be considered.

\begin{theacknowledgments}
B.J. would like to thank the Paul-Urban-Stipendienstiftung for 
supporting her participation in this conference.
\end{theacknowledgments}



\end{document}